\documentclass[conference]{IEEEtran}
\IEEEoverridecommandlockouts
\usepackage{cite}
\usepackage{amsmath,amssymb,amsfonts}
\usepackage{algorithmic}
\usepackage{graphicx}
\usepackage{textcomp}
\usepackage{xcolor}
\usepackage {CJKutf8}
\def\BibTeX{{\rm B\kern-.05em{\sc i\kern-.025em b}\kern-.08em
    T\kern-.1667em\lower.7ex\hbox{E}\kern-.125emX}}
\begin{document}
\begin{CJK}{UTF8}{gbsn}

\title{On the Combination of AI and Wireless Technologies: 3GPP Standardization Progress\\

}

\author{Chen~Sun,~\IEEEmembership{Senior Member,~IEEE,}~Tao~Cui,~Wenqi~Zhang,~\IEEEmembership{Member,~IEEE,}\\
Yingshuang~Bai,~Shuo~Wang~and~Haojin~Li \\ 
Wireless Network Research Department, Sony (China) Limited, Beijing, China \\
\{chen.sun, tao.cui, wenqi.zhang, yingshuang.bai, shuo.wang and haojin.li\}@sony.com}

\maketitle

\begin{abstract}
Combing Artificial Intelligence (AI) and wireless communication technologies has become one of the major technologies trends towards 2030. This includes using AI to improve the efficiency of the wireless transmission and supporting AI deployment with wireless networks. In this article, the latest progress of the Third Generation Partnership Project (3GPP) standards development is introduced. Concentrating on AI model distributed transfer and AI for Beam Management (BM) with wireless network, we introduce the latest studies and explain how the existing standards should be modified to incorporate the results from academia.
\end{abstract}


\section{Introduction}
After 2010, we have witness the big wave of Artificial Intelligence (AI) which surpassing the capability of human in visual recognition and playing complex games. On the other hand, the 4G wireless communication became reality. These two technologies started to merge and enhance each other. On the one hand, the wireless network can support the wide deployment of AI by providing wireless connection during model training and inference for various applications such as autonomous driving, image recognition, etc \cite{Nguyen20}. On the other hand, the AI is applied to solve issues in wireless network such as Beam Management (BM), interference estimation in physical layer, and routing in network layer as well as the resource allocation for mobile operators \cite{Letaief19,wang20}.

Although the work in academia started in the past, the industrialization towards commercial application just started in 2019. In the most prominent standard development of telecommunications, the System Aspects (SA) Technical Specification Group (TSG) of the Third Generation Partnership Project (3GPP) started investigating the impact of Artificial Intelligence and Machine Learning Model Transfer (AMMT) in wireless systems \cite{3GPPTR22.874}. The AI techniques were deployed in the application layer, for example, for image-based object recognition. The 3GPP has continued to investigate the new functions that are needed 1) to support new traffic due to the distribution of AI models, such as, the timely delivery of models for different image recognition tasks and environments \cite{Taylor18}; 2) to support AI model splitting for the distributed deployment of different layers of an AI model \cite{Wang22}; 3) to support the efficient implementation of Federated Learning (FL), such as the use of a 5G Core network (5GC) to provide a list of User Equipments (UEs) with large inter-distances to assist in user selection and reduce training data correlations \cite{sun22}. Such support is realized by enhancing existing functions such as Quality of Service (QoS) management, network information exposure to AI applications, and the allocation of network resources to users selected based on certain criteria, such as location, in FL applications. More aggressive support for user selection in FL can be realized by modifying the backoff window size in the Carrier-Sense Multiple Access (CSMA) mechanism, such that the backoff window size is reduced for selected users to provide them with a greater opportunity to gain access rights to upload local AI models to a server \cite{sun24-ICC}.

On the other hand, “AI for wireless” refers to the use of AI technologies to enhance wireless network functions. Various studies have been carried out to apply AI technologies to different layers of communication protocols \cite{Mao18}. The semiconductor industry has also started to build communication chips with AI capabilities \cite{Ren22}. In 2020, the Radio Access Network (RAN) TSG began investigating the application of AI in RAN functions \cite{3GPPAIMLNR}. The results are expected to guide the design of new communication protocols supporting AI applications. This TSG is studying the following three aspects of this topic: 1) By using AI models deployed on both the Base Station (BS) side, also known as the gNB, and the UE side, the Channel State Information (CSI) can be compressed, transmitted, and predicted. This supports the reduction of the CSI feedback overhead compared with that of the traditional channel estimation and reporting procedure \cite{Guo22}. 2) By deploying AI functions at the UE side, BS side or in the core network side for positioning, the performance in Non-Line-of-Sight (NLoS) scenarios can be enhanced. 3) By applying AI in the BM procedure, the system complexity can be reduced compared with that of the traditional BM procedure.

\section{AI in SA of 3GPP}
\subsection{Enabler for Network Automation (eNA)}
Network automation refers to the utilization of software, often augmented by the advanced functionalities of AI technologies, to undertake a spectrum of tasks. These tasks include planning, deploying, configuring, orchestrating, analyzing, and assuring mobile networks and services. As early as Release 15, 3GPP introduced the network entity of Network Data Analytics Function (NWDAF). This entity functions as an AI black-box, providing network related performance statistics and predictions derived from collecting specific data from one or multiple data sources, e.g. Network Functions (NFs), Operation Administration and Maintenance (OAM) and UE. Subsequent releases have expanded its capabilities to include data collection and network analytics exposure features. These enhancements have been pivotal in augmenting the functionality and usefulness of the feature within the 5GC architecture. Specifically, various types of analytics are now specified, and depending on the Analytics ID(s) requested by a consumer-NF. Probably the earliest implementation on FL in 3GPP is at the eNA, where different NWDAFs can jointly train a model for network analysis \cite{3GPPTS23.288}. The standard also specifies the protocols for NWDAFs form FL process either as server or client.
\subsection{Release 18 AMMT}
In 2019, during the TSG SA \#86 meeting, approval was granted for a work item focusing on the study of traffic characteristics and performance requirements related to AMMT within the 5G System (5GS) \cite{3GPPTR22.874}. The objective is to articulate specific performance criteria, encompassing end-to-end latency, experienced data rate, and communication service availability. Additionally, service requirements, such as AI/ML QoS management, AI/ML model/data distribution and transfer, as well as network performance and resource utilization monitoring/prediction, are being defined. These efforts aim to enable 5GS to effectively support a range of AI/ML operations for diverse applications, including but not limited to image/speech recognition, media editing/enhancements, robot control, and automotive applications.
\begin{itemize}
    \item AI/ML operation splitting between AI/ML endpoints
    \item AI/ML model/data distribution and sharing over 5GS
    \item Distributed/FL over 5GS.
\end{itemize}
The 3GPP SA1 working group accomplished the stage-1 study on AMMT by addressing use cases and potential performance requirements for the support of application layer AI and ML model distribution and transfer within the 5GS. This study also involved identifying traffic characteristics associated with AI/ML model distribution, transfer, and training across various applications. The initiation of the procedures study is to explore capabilities of the 5GS platform to support application layer AI/ML operations, commenced by the 3GPP SA2 working group by the end of 2021.

SA2 has successfully addressed several critical issues, including but not limited to network resource utilization monitoring, safeguarding 5GC information from exposure to UE and third parties, improvements in QoS and policy, and notably, the intricate domain of FL operations. These resolutions contribute to the overall robustness and efficiency of the system, ensuring enhanced privacy, performance, and capabilities in the context of evolving telecommunications standards.

In Release 18 \cite{3GPPTS23.502}, the introduction of FL to the 3GPP marked a significant milestone. The entities primarily involved in contributing to FL are the NWDAF and the AF. The NWDAF plays a central role in conducting model training and updates through a FL server NWDAF, which coordinates with multiple FL client NWDAFs. This collaboration facilitates the aggregation of data from diverse sources while preserving privacy and security. On the other hand, the AF is responsible for scheduling UE participants and exposing the requirements via the NEF to the 5GC. This process adheres to a set of principles, ensuring efficient participation of UEs in the FL process while maintaining network integrity and performance.
\subsection{Device-to-Device (D2D) Aided FL}
For the purpose of offloading tasks from one UE to another in close proximity, many end device vendors have proposed D2D-based or assisted FL. This approach aims to alleviate computation efforts and conserve power by distributing tasks among nearby devices. In this case, relay UEs are necessary for the discovery procedure when selecting FL clients, sometime with the help of AF. Therefore, there is a need to study policy and QoS control, including aggregated QoS thresholds, to support FL over D2D efficiently. 

Unfortunately, it is deemed premature at present to determine the extent to which D2D-based AI/ML functionality could contribute to the system without first deploying a comprehensive AI/ML framework in both the RAN and the 5GC. However, there is an expectation that utilizing D2D aided FL could potentially offer significant benefits in terms of task offloading efficiency and resource optimization once a robust AI/ML framework is established across the network.

\subsection{Release 19 AI/ML}
\begin{figure}[!tp]
    \centering
    \includegraphics[width=1\linewidth]{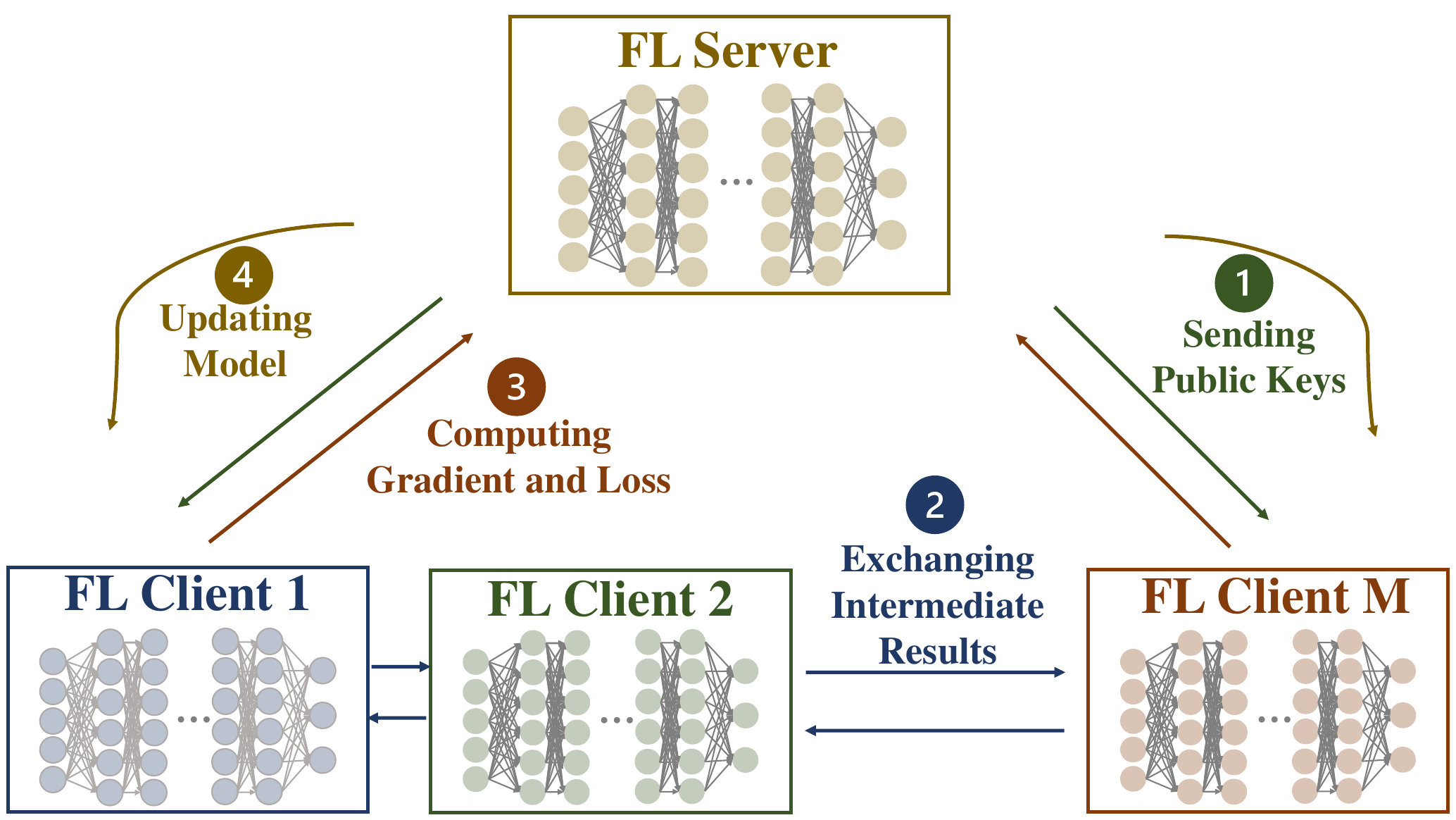}
    \caption{The architecture for VFL system.}
    \label{fig:vfl}
\end{figure}

For the specific objectives of enhancing support for AI/ML within the Release 19 5GC, priority has been given to Location Service (LCS) related enhancements as a pivotal AI/ML cross-domain feature. This prioritization underscores the critical importance of seamlessly integrating AI/ML capabilities into location-based services within the 5GC framework. Central to this effort is the study of a Location Management Function (LMF)-based AI/ML model for positioning, which serves as a key point. This study encompasses various aspects, including how the model is trained, how data is collected, and how inference is performed. Meanwhile, other critical aspects, including UE data collection, model transfer/delivery to the UE, and the alignment of model identification and management, are scheduled for further study and deliberation. These aspects will be addressed in alignment with the decisions and conclusions made during the September 2024 plenary meetings, ensuring comprehensive consideration and alignment with evolving standards and requirements from RAN perspective. The other two prioritized objectives are about supporting Vertical Federated Learning (VFL) and NWDAF-assisted policy control and network abnormal behavior mitigation. For the feature of VFL,  it marks a significant milestone as the group delves into considering data originating from the same sample space but characterized by different feature spaces. An example of a typical architecture regarding VFL system is shown in Fig.\ref{fig:vfl}.

\section{AI in RAN of 3GPP}
3GPP started investigation of AI for wireless since 2018 focusing on BM, position and CSI estimation \cite{3GPPAIMLNR}.
\subsection{Use Cases}

In 3GPP RAN1, there exist three primary AI-based use cases: CSI feedback, BM, and positioning accuracy enhancements \cite{3GPPAIMLNR}. In Release 19, BM and positioning have already progressed to normative work. However, the evaluation results from companies' contributions regarding CSI feedback enhancement have not met expectations. Therefore, participants will further study on how to improve the performance gains.

AI-based BM encompasses two sub-use cases: beam prediction in the spatial domain and beam prediction in the temporal domain. In the spatial domain, a select subset of beams named Set B undergoes measurement, providing input to the model, which subsequently predicts candidate beams for Set A (full set). This approach, when compared to legacy mechanisms, significantly reduces measurement overhead and latency. In the temporal domain, historical measurement data serves as input to the model, enabling the prediction of candidate beams for $K$ future time instances, where $K$ is equal to or greater than $1$. Notably, During certain time intervals, UE does not require measurement, leading to substantial savings in measurement overhead. The aforementioned measured beams can be either wide or narrow, as current standards have not imposed any constraints on it.

AI-based positioning and AI-based CSI feedback. In AI-based positioning, there are two key methods: direct AI/ML positioning for generating exact location coordinates and assisted AI/ML positioning for providing intermediary measurements like timing data. Release 19 focuses on three specific cases of AI positioning, emphasizing the use of models on different sides (UE, NG-RAN node with LMF, and gNB) without compromising user privacy.

For AI-based CSI feedback, two sub-use cases are identified. The first involves using a two-sided model for compressing spatial-frequency domain CSI, reducing report overhead while requiring collaborative model training. The second sub-use case uses a UE-side model for predicting time domain CSI, which decreases RS measurement overhead and allows for L1 reporting similar to traditional methods.

\subsection{Model Management}
Model management encompasses various operations including model activation, deactivation, switching, updating, selection, monitoring, functionality/model identification, and fallback, etc \cite{3GPPAIMLNR}.
\begin{itemize}
\item Model activation/deactivation involves enabling/disabling an AI/ML model for a specific AI/ML-enabled feature.
\item Model update entails the process of updating the parameters and/or structure of a model.
\item Functionality/model identification refers to a process/method of identifying an AI/ML functionality for common understanding between the network and the UE.
\item Fallback occurs when the performance of the current model fails to meet expectations, prompting the UE/network to revert to legacy mechanisms.
\item Model monitoring is a procedure that monitors the inference performance of the AI/ML model.
\item Model switching and selection always execute in the scenarios with multiple models serving a functionality. Model selection is choosing an AI/ML model for activation among multiple models for the same AI/ML-enabled feature. Model switching means deactivating a currently active AI/ML model and activating a different AI/ML model for a specific AI/ML-enabled feature. 
\end{itemize}

In Release 19, there is a necessity to define new signaling and details to facilitate the aforementioned operations. However, several challenges are encountered. For instance, in terms of identification, if functionality identification is employed for model management, the aforementioned operations will be based on functionality. This raises a critical issue: there may be multiple models for one functionality. If the granularity of identification is not detailed enough, some models may remain in an unknown state for UE or network. Consider the following scenario: there are three models utilized for beam prediction at the UE side, operating at speeds of $3$km/h, $60$km/h, and $120$km/h, respectively. However, the network is only aware that the UE can perform beam prediction using AI model and is unaware that the models can be distinguished based on UE movement speed. Consequently, it is unclear for the network to decide on model management operations. However, if model identification is utilized, the aforementioned issue does not need to be addressed, as model management can be based on the model ID. Nonetheless, employing model ID presents its own complexities. For instance, the definition and transmission of model ID can be a complex process, as it may be global or local, physical or logical. Additionally, it cannot be precluded that one model may serve multiple functionalities. As of now, 3GPP has not imposed limitations on identification methods. From the perspective of RAN1, both model identification and functionality identification are deemed acceptable.

For AI-based use cases, it is evident that there are necessary specification impacts to support model management. When the network manages UE-side models, it is essential for both the network and UE to have an aligned understanding of related model information, including model functionality, parameters, and additional conditions. This may involve leveraging UE capability reporting. Regarding additional conditions such as scenarios, sites, and datasets etc., they can be categorized into two groups: network-side additional conditions and UE-side additional conditions. Aligning additional conditions between the network and UE is also significant to ensure consistency between model training and inference. Additionally, these additional conditions can assist in model switching.  when the additional conditions for each model are clearly delineated, facilitating the transitioning to the target model upon the detection of environmental changes becomes more feasible.

Another critical challenge pertains to model monitoring. As so far, model monitoring involves overseeing the performance of the active model, evaluating its suitability for the current scenario. This raises several questions: What are the performance metrics? How should the monitoring procedure be designed? According to research conducted in Release 18, there are several alternatives; for instance, performance metrics may include inference accuracy, data distribution, etc. In Release 19, we may need to make some down-selection among these alternatives. Proponents have also suggested that the monitoring mechanism can be extended to inactive models to assess whether they can be activated. If monitoring is applicable to inactive models, then inactive models have to work to produce predicted results. Consequently, opponents are puzzled by why inactive models are allowed to predict. However, if inactive models cannot work, the execution of model selection becomes uncertain. This raises questions about whether selection should be based on identification information or blind switching. These issues remain unresolved, making everyone more eager for the research results of Release 19. Additionally, employing monitoring to ensure consistency between model training and inference has been proposed. In AI-based use cases, maintaining prediction accuracy by aligning model training and inference with the same scenarios is crucial, and monitoring serves as an effective means to sense scenario changes indirectly. Based on monitoring results, various model management operations can be performed, including model selection, switching, fallback, activation, or deactivation.

\subsection{Performance Evaluation}
In Release 18, three new use cases are introduced in physical layer, the first and significant thing is to evaluate the performance benefits of these new cases is paramount \cite{3GPPAIMLNR}. Several Key Performance Indicators (KPIs) are essential for this assessment, including inference performance, latency, computational complexity, overhead, and hardware requirements. The evaluation methods and metrics for each use case should be tailored to its specific features and requirements.

In AI-based BM use case, system simulation approach is adopted as baseline. The significant KPIs are defined, such as, Top-K/1($\%$) is the percentage of the Top-1 genie-aided beam is one of the Top-K predicted beams. Beam prediction accuracy ($\%$) with $1$dB margin is the percentage of the Top-1 predicted beam whose ideal L1-RSRP is within 1dB of the ideal L1-RSRP of the Top-1 genie-aided beam.

Through the collaborative efforts of various companies, for case 1, it has been observed that by utilizing fixed Set B of beams, which represents approximately one-fourth of Set A of beams, there is a notable enhancement in top-1 DL Tx beam prediction accuracy. The predicted accuracy can achieve between $70\%$ to $90\%$. pecifically when considering the Top-1 DL Tx beam with 1dB margin, it has been demonstrated that accuracy surpasses $90\%$.

When evaluating AI-based positioning, the primary KPIs across all scenarios and use cases revolves around the Cumulative Distribution Function (CDF) percentiles of horizontal accuracy, particularly focusing on $90\%$ (baseline) and optionally including ${50\%, 67\%, 80\%}$. Additionally, reporting of vertical accuracy is optional. In the context of AI/ML-assisted positioning, the output of AI/ML models can encompass various types of information, such as Time of Arrival (ToA), Received Signal Strength Difference (RSTD), Angle of Departure (AoD), Angle of Arrival (AoA), LoS/NLoS indicators, among others.

The inference accuracy in AI-based positioning is notably influenced by the channel environment, particularly in NLoS scenarios. Based on companies' contributions that for foundational performance without generalization, AI/ML models undergo training and testing using datasets from the same deployment scenario. AI/ML-based positioning demonstrates significant enhancements in positioning accuracy compared to existing Radio Access Technology (RAT)-dependent positioning methods. For instance, in an InF-DH scenario with clutter parameter settings of ${60\%, 6m, 2m}$, AI/ML-based positioning achieves horizontal positioning accuracy of smaller than $1$m at CDF equals to $90\%$, in contrast to \textgreater $15$m for conventional positioning methods.

Regarding the performance evaluation of AI/ML-based CSI feedback enhancement, system simulation serves as a baseline. KPIs and metrics include:
\begin{itemize}
\item The evaluation of CSI compression includes key performance indicators such as Squared Generalized Cosine Similarity (SGCS) and/or Normalized Mean Square Error (NMSE) for assessing the accuracy of CSI outputs generated by AI/ML. SGCS is computed separately for each layer, when rank$> 1$.
\item Evaluation of CSI prediction involves calculating intermediate KPIs for each predicted instance when the AI/ML model produces multiple predictions.
\end{itemize} 
In \cite{3GPPAIMLNR}, simulation results for CSI compression from various companies are presented. For example, in deployment scenarios where scenario A is UMi and scenario B is UMa, scenario A is UMa and scenario B is UMi, or scenario A is UMa and scenario B is InH, performance gains ranging from $-1.69\%$ to $-31.6\%$ degradation are observed. As for the CSI prediction, If UE speed B is either $30$km/h, $60$km/h or $120$km/h, or if UE speed B is $10$km/h and UE speed A is either $60$km/h or $120$km/h, moderate to significant performance degradations ($-2.01\%$ to $-76.85\%$ loss) are experienced. 

Based on the observations mentioned above, the performance evaluation results for CSI compression and CSI prediction were highly unsatisfactory. Therefore, in Release 19, the normative work were not pursued; instead, the focus remained on maintaining the status of study items.

\section{From academia to standards}
\subsection{AI Model Distributed Transfer}
\subsubsection{D2D Aided FL}
With the booming growth of terminal intelligent services, the demand for wireless network support for intelligent models is becoming increasingly high. Generally, single UE is difficult to support the training of ML models with higher accuracy and stronger generalization. In addition, there is data isolation problem since UEs are unwilling to share their own datasets due to information privacy and security issues. Therefore, 3GPP RAN1 and RAN2 work items consider the AI/ML model transfer between gNB and UEs as a potential topic. The distributed framework of FL adopts the requirements of AI/ML model transfer properly. In particular, FL empowers multi-party collaboration while protecting the information privacy and security. However, during the FL training process, there still exists three bottlenecks, which are communication quality enhance, computation and memory resource reduction and communication resource reduction. Along this line, our works utilize D2D to alleviate the three bottlenecks of FL.

\textbf{Communication Quality Enhance:} In some circumstances, the specific UE or a group of UEs under abysmal communication quality and hard to support the local model uploading. Based on this scenario, the typical UE transmits its local model to the near UE through the D2D link and the near UE aggregates its own local model and typical UE's local model. The aggregated model transferred to the gNB to solve the problem of unreliable communication. Different from the typical FL, the UEs need to transmits the D2D link information to gNB. Furthermore, whether the D2D link transmission was successful should be notified to the gNB or the other UEs.

\textbf{Computation and Memory Resource Reduction:} Considering the scenario that UEs cannot train the complete model with limited computation and memory resource. Based on this scenario, UE establishes D2D link with its peripheral devices, which are defined as Auxiliary Equipment (AE). According to the computation and memory capability of the typical UE and AE, the complete model is split into two parts by Split Learning (SL). Through sharing part of the training pressure with AE to train the complete model sufficiently with limited UE computation and memory capability. During the aforementioned process, gNB needs to determines the split point of complete model and the communication resource allocation of uplink, D2D link and downlink.

\textbf{Communication Resource Reduction:} To reduce the communication resource, our work simultaneously considers UE selection, UE execution order selection, and model transmission link selection based on channel state of D2D between UEs and other possible links (such as, relay chain) are significant factors to make decisions. In addition, UE content, which means that whether the UE agrees with the link options available for model transmission, is added to protect the information privacy. It is worth noting that the UE content maybe a potential signaling for 3GPP. In \cite{SUN24SPAWC}, a UE transmits on the difference between its local model and the model overheard from nearby UEs, thus reducing the uplink traffic in each loop of FL. These technologies require the network to select users based on their D2D channel qualities.

\subsubsection{Knowledge Distillation}
One of the most critical challenges in deploying AI models on end devices is model compression, owing to constraints in computation and memory. When considering the entire wireless communication system, transmitting a billion-parameter model over the air interface becomes unrealistic, particularly amid uncertainties in the wireless channel stemming from environmental fluctuations and increasing user numbers. In such scenarios, leveraging knowledge distillation becomes invaluable for resizing the model to better fit within end devices.

Ongoing discussions within the 3GPP, specifically within the RAN working groups RAN1 and RAN2, are focused on model delivery and transfer. There is a debate on whether a standardized solution is necessary for this process. Delivering an AI/ML model over the air interface, whether it's the parameters of a model structure or a new model with parameters, could be seen as transmitting a large packet of data that remains transparent at the signaling level to the UE. However, it's important to note that for use cases requiring high efficiency in real time, there are potential benefits to incorporating RAN-level parameters related to AI/ML models in the process of model updates and iteration.

Our work \cite{FL-dist} introduced the model compression through knowledge distillation to alleviate network requirements in model distribution. Essentially, the methods efficiently distill knowledge from large models trained on centralized data to smaller models trained in a federated manner across distributed nodes. By compressing the models, our approach aims to diminish communication overhead and enhance the efficiency of FL, especially in resource-constrained environments.

\subsection{AI for Beam Management}
One of the topics discussed regarding to the physical layer in 3GPP is AI for BM which is critical to the 5GS due the usage of mmwave. Various studied have been carried out in academia \cite{Khan23}. 

\subsubsection{Model Input and Output}
An apparent task for applying an AI model for BM is defining the input and output of the AI based BM system. In \cite{Xue21}, we attempt to reduce number of beam switching for a moving UE. The training data is collected by selecting the beam that satisfies the performance requirement while having the highest dwelling time for the moving UE. Using this data we can train an AI model that output the beam with maximum dwelling time for a given location. This requires that the standards defines a message protocol for UEs report their location and even moving speed and direction. the BM complexity by Depending on how the input data is constructed the prediction could be based on different criteria. In \cite{Make22}, deep learning is applied to predict a few beams with the sum probability larger than a given threshold. Then, the traditional beam measurement complexity is reduced since only a subset of beam shall be measured. AI is also used to predicted beams with high signal quality or longer duration time. Correspondingly，the standard defines criteria of beam prediction together with the prediction results as output. 
mmWave MIMO systems usually requires separate digital and analog precoders, and the complexity increases with the number of antennas.
In \cite{Chai20GC}, the authors exploited DNN to predict precoding matrices, using imperfect CSI. In addition, the model is pruned to both reduce the model size and improve the processing speed almost without degrading the accuracy.

\subsubsection{Model Management}
Having a universal AI model trained for various propagation environments is infeasible. AI model selection based on propagation environment is proposed in \cite{Sun-24-OJCOM}. Classification of environment is based on the measurement of the BS downlink beams. Different AI models are selected according to the classification result. This requires the system to report on multiple beams measurement results rather than just reporting the best downlink beam. Furthermore, AI based performance monitoring is proposed. Based on the past history data on the performance of both traditional and AI based BM. The AI model can predict the performance of the two approaches and select either one based on the prediction result.

\subsubsection{Data Collection}
For the beam prediction in time domain, the historical measurement data is collected using for predicting the candidate beams, it is essential to investigate the optimal time window size for collecting input data to minimize unnecessary measurement overhead. \cite{10334007} propose that the size of the time window for collecting input data should be determined based on the UE speed. This is because the characteristics of the time-domain channel are vital in deciding the ideal time window size for model inputs. A key factor in evaluating these characteristics is the UE speed. Higher speeds in UE lead to a fast-fading channel state, whereas lower speeds result in a slow-fading state. The channel's coherent time is inversely proportional to the UE speed, meaning that a fast-fading channel exhibits a shorter coherent time than a slow-fading channel. For enhanced performance of the AI model, it is crucial that there is a strong correlation between input and output data. Therefore, the UE movement speed significantly influences the optimal time window size for data collection in AI models.

\section{Conclusion}
In this article, we introduced the progress in industry regarding to the combination of the AI and wireless communication network. We introduced the latest achievements in academia related to FL and BM. In particular we emphasized on how the system shall be designed to supprt a better combination of AI and wireless.

\bibliographystyle{IEEEtran}
\bibliography{ref}

\end{CJK}
\end{document}